\documentclass[preprint,aps,bibnotes,endfloats]{revtex4}

\usepackage{amssymb}
\usepackage{graphicx}
\usepackage{multirow}
\usepackage{dcolumn}
\usepackage{bm}

\usepackage{graphicx}

\begin{document}

\title{Molecular dynamics simulation of the recrystallization of amorphous Si layers: Comprehensive study of the dependence of the recrystallization velocity on the interatomic potential}
\author{C. Krzeminski, Q. Brulin, V. Cuny, E. Lecat, E. Lampin \footnote{evelyne.lampin@isen.iemn.univ-lille1.fr} and F. Cleri
} \affiliation{IEMN/ISEN, UMR CNRS 8520, Avenue Poincar\'e, 59652 Villeneuve d'Ascq Cedex, France
\\}

\vspace{10 cm}

\begin{abstract}
The molecular dynamics method is applied to simulate the recrystallization of an amorphous / crystalline silicon interface. The atomic structure of the amorphous material is constructed with the method of Wooten, Winer and Weaire. The amorphous on crystalline stack is afterwards annealed on a wide range of temperature and time using five different interatomic potentials : Stillinger-Weber, Tersoff, EDIP, SW115 and Lenosky. The simulations are exploited to systematically extract the recrystallization velocity. A strong dependency of the results on the interatomic potential is evidenced and explained by the capability of some potentials (Tersoff and SW115) to correctly handle the amorphous structure while other potentials (Stillinger-Weber, EDIP and Lenosky) lead to the melting of the amorphous. Consequently, the interatomic potentials are classified according to their ability to simulate the solid or the liquid phase epitaxy.
 
\end{abstract}


\maketitle

%

\section{Introduction}

The pre-amorphization of silicon by high dose ion implantation is one of the identified processes for the realization of the ultra-thin junctions of the technological nodes 65 nm and below of the silicon microelectronics industry. The transformation of crystalline (c-Si) into amorphous silicon (a-Si) on a layer close to the surface reduces the channeling of the dopant consecutively implanted. Such a technological stage is therefore required to limit the penetration into the substrate of small radius atoms such as boron, the standard $p$ dopant in microelectronics \cite{Tsaur83,Seidel85}. The pre-amorphization and the dopant implant might be followed:

\begin{itemize}
\item by an annealing at a temperature low enough ($<$ 700 $^{\circ}$C) to recrystallize the amorphous layer and activate the doping, but insufficient to cause a significative diffusion of the dopant. Reference \onlinecite{Timans04} gives such an example, with a depth of the junction at 10$^{18}$ cm$^{-3}$ of $[$16-19$]$ nm compatible with the values $[$10-19$]$ nm targeted by the ITRS roadmap \cite{ITRS}. Likewise, the high level of activation allows to reach sheet resistances of $[$405 -785$]$ $\Omega/\square$ to compare with the range $[$760-830$]$  $\Omega/\square$ of the ITRS Roadmap. 
\item by an annealing at very high temperature (laser or spike annealing). In the case where the melting temperature of a-Si is reached, the profile after annealing has an ideal ``square-like" shape as the diffusivity of the dopant is higher in the liquid than in the crystal. An example of such a profile is given in the reference \onlinecite{Hernandez03}.
While ideal for a full wafer implantation, this solution however is not always compatible with the other technological stages \cite{Lindsay03}. The other alternative is to anneal just beneath the melting temperature (ex : 1300 $^{\circ}$C) during a very short duration (ex : spike of 330 $\mu$s wide at 1250 $^{\circ}$C) to limit the dopant diffusion \cite{Fiory99}.
\end{itemize}

In the two cases, it is critical to have a good understanding of the recrystallization process of the amorphous layer to develop such technologies. If, on the one hand, the recrystallization velocity by solid phase epitaxy (SPE) of an amorphous layer is well known from experiments in the intrinsic case \cite{Olson88}, this is not the case in presence of impurities where accelerations, slowdown, compensation and even hysteresis are observed \cite{Olson88}. Moreover in this case, the final distribution of the dopant is intimately related to the velocity of recrystallization. Now, this case of doped a-Si is precisely of current interest for the formation of junctions. In the absence of an experimentally established law, is it possible to calculate at the atomic level the recyrstallisation velocity of doped a-Si ? Atomistic simulations of a-Si, of its interface with c-Si and of its recrystallization have already been published \cite{Caturla94,Caturla96,Bernstein00,Motooka00,Motooka01,Humbird02,Marques03,Mattoni04}. However in these works the atomistic simulations of the recrystallization of a-Si are not always exploited to extract the recrystallization velocity $v_\mathrm{cryst}$. The comparison with experiments is most often limited to the activation energy of $v_\mathrm{cryst}$. Moreover, the description of silicon by means of empirical interatomic potential is not unique and the question of the most adapted one(s) for this particular case is still open. The aim of this paper is to present a systematic study of the recrystallization velocity as a function of the annealing temperature for the various interatomic potentials available for silicon. The present work has been devoted to the intrinsic case, that already appears an issue by itself. 

In this paper, we begin by a presentation of the method of simulation we have used to model the a-Si, its recrystallization and the extraction of the recrystallization velocity. The results are then presented for five different interatomic potentials in the second part. In the last part, the wide differences obtained as a function of the interatomic potential are analyzed and an explanation is proposed based on a possible transformation of the amorphous layer into a liquid one.

\section{Method}

This section is first devoted to a discussion about the different empirical interatomic potentials used in the molecular dynamics simulation. Next, the description of the method  used to build the a-Si cluster and to its properties is presented. The generation of the a-Si/c-Si structure is also reported. Finally, the simulation of the recrystallization of a-Si on c-Si with molecular dynamics  is described. The systematic exploitation of these atomistic simulations to determine the recrystallization velocity is also explained in details. 

\subsection{Molecular dynamics using interatomic potentials}

In this article, the annealing of the a-Si on c-Si stack was investigated by means of molecular dynamics simulation employing empirical interatomic potentials. Interatomic potentials are an efficient way to describe the interaction between atoms using a continuous and derivable functional of the atomic positions. As they do not contain deep physical effects as in ab-initio description, their domain of validity is not universal. Most often, interatomic potential are developed on a particular application such as the study of surfaces or the description of a phase transition. Five empirical potentials developed for silicon were tested in this study : the Stillinger-Weber potential in its original parametrisation \cite{SW} and in a recent parametrisation called SW115 \cite{SW115} designed for a better description of the melting of a-Si and c-Si, the third formulation of the Tersoff's potential \cite{Tersoff}, the Environment Dependent Interatomic Potential (EDIP) \cite{EDIP} and the Lenosky's pontential \cite{Lenosky}. To our knowledge, these are the main potentials used in the literature to describe silicon. The 4 functional forms (Stillinger-Weber and SW115 have the same formulation) contain an attractive and a repulsive part given in the form of continuous functions of the relative distance between pairs of atoms, with a two-body and a three-body character, respectively. These functional forms are adjusted to reproduce the $T=0$ equilibrium crystalline structure, cohesive energy, elastic constants and possibly other structural or energetic parameters of crystalline Si. The Stillinger-Weber potential has a three-body term that imposes the angle $\theta$ between the three neighbors to $\cos\theta = -1/3$ and therefore strictly enforces the diamond structure. The EDIP potential is inspired from the Stillinger-Weber potential, but it adapts to the coordination (4, 3, 2,...) of the atom and it is therefore more general. The formulation of the Tersoff potential is slightly different from the two previous ones, and has the form of a pair potential with a bonding strength that contains the dependence upon the local atomic environment. Finally, the Lenosky is a modified embedded-atom method including a pairwise embedding functional and an angular term. Which potential should be more adapted to our case is not obvious a priori. Therefore simulations with these 5 different potentials were performed.

\subsection{Amorphous cluster}

Amorphous silicon in silicon devices is created by ion implantation of heavy ions. Ideally, the best approach would be to mimic the experiment and therefore to simulate the ion implantation. However, it is not currently realistic since the simulation of the implantation of an ion and of all the interactions and collisions with the atoms of the target by molecular dynamics is already computer resources consuming, and many events of that kinds should be simulated to obtain an a-Si layer of good quality. The other alternative is to construct a cluster of amorphous silicon and to glue it on crystalline silicon. The main two methods to form an amorphous structure starting from the crystal are a) the disordering of tetrahedral interatomic bonds by the method of Wooten, Winer and Weaire (WWW)\cite{Wooten85} and its derivatives \cite{Barkema00,Valiquette03} and b) the simulation of an idealized a-Si, obtained by a melting followed by a quenching to a solid but amorphous state \cite{Luedtke89,Ishimaru97}. We will call this second method the ``melting-quenching". Each of these two methods has its advantages and drawbacks. The main disadvantage of the WWW method is to enforce a tetravalent coordination, while it is experimentally known \cite{Coffa93} that a-Si contains defects which may be floating or dangling bonds. The simulation of the melting-quenching on the other hand, done using a molecular dynamics simulation with an empirical potential, relies first on the ability of the potential to correctly describes a liquid at elevated temperature, and second to correctly describe the amorphous structure upon quenching, which is not unconditionally ensured for neither of these two aspects as emphasized later (see Section \ref{Discussion}).

In the present work, the results are obtained with a WWW a-Si. The reason for this choice is first to construct an amorphous cluster of good quality, and next paragraph will discuss of this point. Moreover, the objective was to test which of the numerous empirical interatomic potential developed to model silicon \cite{SW,Tersoff,EDIP,SW115,Lenosky} is the more adapted to keep the amorphous structure and to give the recrystallization velocity measured by solid phase epitaxy experiments \cite{Olson88}.  Using the melting-quenching method would have required to construct for each potential its corresponding amorphous cluster, which, while more rigorous than the use of a unique WWW amorphous cluster, would have introduced different amorphous structures. Therefore all the results presented in the section \ref{SectRes} have been obtained with the same initial a-Si cluster constructed by the method presented just below. On the other side, we believe that, once the choice of the most adapted interactomic potential has been made, the better approach should on the contrary be the simulation of the melting-quenching-recrystallization with this potential.

\subsubsection{Construction}
The WWW \cite{Wooten85} a-Si structure is obtained using an algorithm of bond switching between pairs of first neighbours. The total energy associated to this bond reordering is originally calculated using a Keating potential \cite{Keating}. A Maxwell-Boltzmann probability is then determined with this energy and a  temperature. In this way, an elevated value of the temperature can be used at the beginning of the amorphisation to strongly disorder the structure, and be subsequently decreased to reduce the number of defects in order to obtain a very realistic a-Si. This procedure was applied to an initial cluster of 4096 atoms of c-Si. 

\subsubsection{Properties}

The current paragraph presents the properties of the WWW a-Si cluster of 4096 atoms and the comparison with published experiments. 

The side of the initial cubic cluster of c-Si is of $L_\mathrm{c-Si}$=43.44 \AA. The density before amorphisation is therefore equal to $d_\mathrm{c-Si}$=$4096/L_\mathrm{c-Si}$ = 0.0500 at/\AA$^3$. After amorphisation, the dimensions of the cluster are slightly reduced: $L_\mathrm{a-Si}$=43.38 \AA. The density is therefore 0.4\% higher ($d_\mathrm{c-Si}$=0.0502 at/\AA$^3$). This feature is in disagreement with experiments where a reduction of the density
    of approximately 1.8 \% has been measured \cite{Custer94}. This difference might be due to the dilatation and to the strain induced by the implantation at high dose and high energy ($>$ MeV) used in the experimental articles \cite{Volkert91}, both effects neglected in the model. It is not obvious to simulate these macroscopic effects at the atomic level. Moreover, it is not sure that this rather slight difference is significative of a problem in the model.

The angular distribution of the bonds between first neighbours is shown in Fig. \ref{FigAngDist}. The mean angle of the distribution is equal to 109.1$^\circ$ and the standard deviation is equal to 10.6$^\circ$, in excellent agreement with the work of Djordevic et al.\cite{Djodjevic95}.

The pair correlation function $g(r)$, i.e. the probability to find an atom between the spheres of radius $r$ and $r+dr$ centered on any atom is given in Fig. \ref{FigRadDist}. $g(r)$ is calculated using the radial distribution $G(r)$~:
\begin{equation}
g(r)=\frac{G(r)}{4\pi r^2 \rho_{0}}
\label{EqPCF}
\end{equation}
where $\rho_{0}$ is the density of c-Si. The radial distribution is the number of neighbours at the distance $r$~:
\begin{equation}
G(r)=\frac{1}{N}\sum_{i,j} \delta(r-|R_{i}-R_{j}|)
\label{EqRadDis}
\end{equation}
where $|R_{i}-R_{j}|$ is the distance between atoms $i$ and $j$ and $N$ the total number of atoms. The abscissa of the maxima of the pair correlation function (see Fig. \ref{FigRadDist}) corresponds to the distances between first (2.35 \r{A}), second (3.83 \r{A}), third (5.43 \r{A}) and even fourth (6.62 \r{A}) neighbours. The peaks are of course enlarged due to the disorder in a-Si compared to c-Si, but at least three peaks are clearly distinguishable and the first and second peaks are clearly separated, with a gap to $g(r)$=0 between the two peaks. The pair correlation function is not directly comparable to experiments. Therefore its Fourier transform was calculated to extract the structure factor $S(Q)$, given in Fig. \ref{FigFactStruc}. The shape, the position and the intensity of the peaks is in perfect agreement with the structure factor measured by Laaziri et al. \cite{Laaziri99,Laaziri99_2} on a-Si obtained by ion implantation, except a slight underestimation of the intensity of the first peak. This agreement is significative of the good description of the local order in the WWW a-Si. 

To conclude, although the discrepancy on the density, the WWW a-Si has excellent geometrical characteristics and the structure factor is in good agreement with experiments. Therefore the quality of the a-Si structure is adapted to the objective of this study.

\subsection{Construction and properties of the a-Si/c-Si structure}

The resulting amorphous cluster of 4096 atoms (43.38 \AA~ each side) was then truncated  and glued along the [100] surface  of the c-Si cell to form the interfacial structure \cite{Mattoni04, Brambilla00, Motooka00}. The  structure contains  1905 atoms and the crystal part  of the interfacial structure  corresponds to  50\% of the global thickness. It has been observed that the crystalline thickness  is an important factor since it appears in some cases that a recrystallization of the a-Si layer was obtained for the bigger structure while the disorder propagated in the c-Si part of the smaller structure (21\% of the global thickness)  although the two atomic layers at the bottom of the structure were frozen at their perfect diamond lattice site to eventually compensate the thin thickness of the crystalline part. Tests on  bigger stack (8034 atoms) did not give significantly different recrystallization velocities than with the stack of 1905 atoms. Since the computational time is drastically increased for this stack of 8034 atoms, we dropped it and focused the work on the 1905 atoms interfacial system. Table \ref{table:interface_coordination} presents the mean coordination number and the silicion defects distribution of the initial interface. The statistics have been estimated for all the silicon atoms located at less than 2.35 \r{A} of the  interface both for the cristal and the amorphous part. A cut-off distance of 2.7 \r{A} has been used to calculate the first neighbours. Large amount of threefold and twofold silicons defects are present initially due to the difficulty to perfectly align an amorphous interface and a perfect crystal. 

In order to improve the interface quality, prior to the main recrystallisation step, an equilibration step has been  assumed. An objective is  to relax the interface  and to reach the temperature targeted value for the main recrystallisation simulation.  Periodic boundary conditions only along the lateral directions have been used allowing the system  to relax along the regrowth direction.  The equilibration phase is performed during 4 ps at the targeted temperature value using a 2 fs time step for the velocity Verlet algorithm.  During this equilibration, the temperature is imposed by the scaling of the velocity which ensures an efficient convergence of the temperature as shown in Fig. \ref{FigTemp}. The empirical atomic potential used in this  pre-annealed step is identical to the one used for the recrystallisation simulation. Table \ref{table:interface_coordination} presents the evolution of the mean coordination number distribution for the different interatomic potentials. First of all, it has been checked  that during this equilibration step, no real recrystallisation occurs (the interface variation  is less than 0.4 \r{A} for all the  inter-atomic potential). The equilibration step improves the quality of the interface since the fraction of fourfold silicon atoms is  enhanced for almost all interatomic potentials. On the other side, the choice of the interatomic potential has a clear influence on the defects at the interface even for a very short annealing. After equilibration, both Tersoff and SW115 tend to favorize the presence of threefold defects. On the other side, fivefold silicon atoms are the most dominant defects  with the SW, EDIP and Lenosky potentials.

\subsection{Recrystallization of the amorphous layer}

The recrystallization of the amorphous layer has been simulated  after the equilibration step in an second step where a Nos\'e-Hoover thermostat \cite{Nose-Hoover} is applied. The acquisition begins  during a number of time steps sufficient to observe the recrystallization of all or of one a part of the a-Si layer. During this phase, the positions of the atoms are saved to allow a post-treatment explained in the following part. The interval between two backups is adapted to the speed of the recrystallization : from 10 ps to 1000 ps. Once the simulation has ended, i. e. the amorphous layer is partially or totally crystallized, the positions stored during the simulation are exploited to extract the recrystallization velocity. Several methods have been used in the literature to quantify the motion of the recrystallization front. Bernstein et al. \cite{Bernstein00} classify the atoms as ``crystalline" or ``amorphous" on the basis of a geometric criterion called $\chi$ and equal to the sum of the absolute values of the dot products between bond directions around each atom and the bond directions in the underlying crystal lattice. This parameter $\chi$ is equal to 4 in the crystal and 2.7 in the a-Si. The velocity of recrystallization is extracted from the number of crystal atom versus time. Marqu\'es et al. \cite{Marques03} extract the velocity in atom/ps from the slope of the potential energy curves, since the recrystallization is accompanied by a decrease of this energy. Motooka et al. \cite{Motooka00} only specify that the a-Si/c-Si interface is defined as a middle of their rough interface, without additional information on a criteria used. Albenze et al. \cite{Albenze04} define an order parameter equal to 1 in solid, 0.7-0.8 in an amorphous state and 0.2-0.4 in liquid, from the angles between each atom and its first neighbours. Next, the sample is divided into slices, and the parameter is calculated inside each slice which is consecutively classified into one of the three states. This method helps to follow the displacement of a crystalline/amorphous interface. In our opinion, the most rigorous analysis is made by Mattoni and Colombo \cite{Mattoni04}. They calculate a structure factor along the growth direction $S(z)$ that simply quantify the crystalline order in 2D slices in the $z$ direction. This structure factor is equal to 1 when the plane is ordered (case of a crystal at 0 K) and around 0 when the order is lost. A detection of $S(z)$ = 0.5 is used to easily place the a-Si/c-Si interface and further deduce the recrystallization velocity. The same criterion has been used to extract the recrystallization velocity from our simulations. Fig. \ref{FigSz} gives an example of structure factor for a stack of 8034 atoms. The structure factor along the growth direction is aligned with the representation of the atoms to emphasize the link between order/disorder and the value of $S(z)$. The interface position obtained at $S(z)$ = 0.5 is reported versus time in Fig. \ref{FigFitInterface}. Until $\approx$ 240 ps, the position increases linearly with time : the disordered layer recrystallizes at a roughly constant velocity. After 240 ps, the position saturates, the whole layer has recrystallized. The velocity of recrystallization is estimated from the fit of the linear region (see Fig. \ref{FigFitInterface}). In this particular case, a value of 0.015 nm/ps was found. This procedure is applied to 50 to 100 files of 1905 atomic positions, for 4 to 6 temperatures for each of the 5 potentials. The evolution of the a-Si on c-Si was studied up to $\approx$ 100 ns for the slowest cases, a computer job of two months. 

\section{Results and comparison with published simulations}
\label{SectRes}

The results of our molecular dynamics simulation of the recrystallization of an a-Si on c-Si stack are presented in Fig. \ref{FigVelCalc} for the Stillinger-Weber, Tersoff, EDIP, SW115 and Lenosky potential. This figure evidences a strong dependence of the velocity of recrystallization on the interatomic potential. At 1200 $^{\circ}$C, there is for example 4 orders of magnitude separating the results with the Stillinger-Weber potential from the results with the Tersoff potential. These differences might also be even more pronounced at low temperature, but the limit of our simulations at time being is 2$\times$10$^{-6}$ nm/ps (one monolayer recrystallized after 100 ns) and does not allow to get additional points for the Tersoff potential.

Very few data are available in the literature to perform a comparison with our calculations. As for the Stillinger-Weber potential, the only point to our knowledge can be deduced from Fig. 1 of Mattoni and Colombo \cite{Mattoni04} : at 1200K, we have estimated a velocity of $\approx$ 2$\times$10$^{-3}$ nm/ps. However, their calculation was made i) on a system obtained by melting-quenching with the Stillinger-Weber potential, although this procedure is known to give an amorphous layer of poor quality \cite{Broughton87} and ii) including B atoms in the Si lattice to study their clustering upon annealing. It is known that dopant atoms modify the velocity of the recrystallization \cite{Olson88}. The direct comparison with our calculations would therefore not be conclusive, although the order of magnitude is roughly comparable (around 10$^{-2}$ nm/ps in our case versus the value of 2$\times$10$^{-3}$ nm/ps in the paper of Mattoni and Colombo \cite{Mattoni04}). 

Concerning the simulation of the recrystallization with the Tersoff potential, the velocities calculated by Marqu\'es et al. \cite{Marques03} in atom/ps is not useful for the comparison with our simulations. Motooka et al. \cite{Motooka00} give the velocity as a function of the inverse temperature. Their results are reported in Fig. \ref{FigCompSimu}. Their simulations with an a-Si system obtained by melting-quenching give two regimes, a low temperature regime with an activation energy of 2.6 eV and a high temperature regime with an activation energy of 1.2 eV. On the contrary, experiments give an Arrhenius behavior from 470 to 1350 $^{\circ}$C with a single activation energy of 2.68 eV \cite{Olson88}. We do not observe these two regimes, but a single one with an activation energy of 2.99 eV. Nevertheless, our results are in perfect agreement with the results of Motooka et al. in the low temperature regime (see Fig. \ref{FigCompSimu}).

Bernstein et al. \cite{Bernstein00} have published a figure of the recrystallization velocity as a function of the inverse temperature between 700 and 1100 K for the EDIP potential. They also obtained two regimes. The low temperature regime ( $<$ 950 K) has a very low energy of activation (0.4 eV) \cite{Bernstein98}. The regime at high temperature has an activation energy of $\approx$ 2.0 eV. The fit of their points in this regime is given in  Fig. \ref{FigCompSimu} together with our results. Our result is 3 times higher at 1100 K  and the difference increases when the temperature decreases. However a  direct comparison is difficult since different methods have been used. For example,  the velocity of recrystallization is calculated by identification of the ``crystal" and ``amorphous" atoms, not with a structure factor. Moreover the amorphous part of the atomic structure is generated by a procedure of melting-quenching.

To our knowledge, there are no published values for the SW115 and Lenosky potentials. 

To conclude, the comparison between our simulations and the literature is difficult. Many differences exist in the method to generate the amorphous layer or to estimate the interface position. However it can be noticed a relatively good agreement with the simulations of Motooka et al. \cite{Motooka00} for the low temperature regime. 

\section{Discussion}
\label{Discussion}

Although the results of the simulations presented in Fig. \ref{FigVelCalc} seems at first glance very disparate, a deeper analysis allows to regroup them under two categories. The first category is made up of the results obtained with the Tersoff and SW115 potentials. Both interatomic potentials give an Arrhenius behavior and smaller values of the velocity of recrystallization. The second category includes the Stillinger-Weber, the EDIP and the Lenosky potential. Using these three potentials, the velocity of recrystallization does not follow an Arrhenius law any more but presents a maximum followed by a decrease of the velocity at high temperature. Moreover, the values of the velocities are higher in this second case. In this section, we firstly compare the results of each category with the measurements of the recrystallization velocity in the case of a SPE and LPE (Liquid Phase Epitaxy) respectively, before presenting an analysis of the order parameter and melting temperatures to evidence SPE in the first category and LPE in the second.

\subsection{Comparison with experiments}
The first category of results, obtained with the Tersoff and SW115 potentials, is characterized by an Arrhenius behavior similarly to the numerous measurements of the velocity of recrystallization of a-Si on c-Si reviewed by Olson and Roth \cite{Olson88}. Fig. \ref{FigFitTersoffSW115} presents the experimental results of Olson and Roth, our simulations with the Tersoff and SW115 potential and their fits. The activation energies are 2.99 eV for Tersoff and 1.87 eV for SW115. Not only the activation energy obtained with the Tersoff potential is close to the value of 2.68 eV obtained by measurement, but also the absolute value of the velocity. As for the SW115, the behavior is globally similar, although the recrystallization is accelerated in comparison to the experiments, especially at low temperature (lower activation energy).

The second category includes the Stillinger-Weber, the EDIP and the Lenosky potentials. These three potentials have in common higher values of the velocities, up to 0.02 nm/ps i.e. 20 m/s. These results are orders of magnitudes far from the law of Olson and Roth \cite{Olson88}, they don't seem to have anything to do with the phenomenon measured by these authors. However, velocities of m/s are typical of the recrystallization of silicon melted by pulse laser \cite{Galvin85} and of explosive recrystallization of a-Si during pulsed laser irradiation \cite{Thompson84}. In both cases, the speed of the recrystallization is governed by the transformation of liquid silicon (l-Si) into c-Si. In the case of explosive recrystallization of a-Si, the liquid is formed at the a-Si/c-Si interface due to a local heating by the release of heat generated by the transition from a-Si to c-Si \cite{Thompson84}. The speed of crystallisation of l-Si has been measured by Galvin et al. \cite{Galvin85}. These measurements are presented in Fig. \ref{FigSWExpFit} together with the results obtained with the Stillinger-Weber potential. The simulations at high temperature are in excellent agreement with these measurements. Moreover, the transition-state theory of crystal growth gives the following expression for the velocity of recrystallization \cite{Kluge89}:
\begin{equation}
v(T)= C\exp \left(-\frac{Q}{kT}\right)\left(1-\exp\left(-\frac{L(T_0-T)}{kT}\right)\right)
\label{EqVelLiqu}
\end{equation}
where $Q$ is the activation energy for viscous or diffusive motion in the liquid, $L$ is the heat of fusion of the phase transformation, $T_0$ is the equilibrium melting temperature and $C$ is a constant. Our simulations were fitted to this law, and Fig.  \ref{FigSWExpFit} shows that an excellent adjustment can be obtained with $T_0$ = 1600 K and $Q=$ 0.54 eV. The results obtained with the EDIP and Lenosky potentials can also be perfectly calibrated with Eq. \ref{EqVelLiqu}, but the agreement with experiments is less good in these two cases : EDIP gives a slower recrystallization, and Lenosky is shifted towards smaller temperatures.

The conclusion is therefore that the recrystallization velocities obtained with the Stillinger-Weber, EDIP and Lenosky potentials are very different of the results obtained for a solid phase epitaxy, but on the other hand they are typical from the solidification of a liquid. The question is now on the origin of this behavior, since the initial system is indeed an a-Si / c-Si stack, and not a l-Si / c-Si stack. This problem is studied in the following section.

\subsection{Analysis : SPE versus LPE}

In order to analyze in more details the structure of the initial a-Si layer upon annealing, we have calculated the order parameter defined by Albenze et al. \cite{Albenze04}. First the parameter called $A$ is calculated for each atom:
\begin{equation}
A=\sum_i \left(\cos\theta_i + \frac{1}{3} \right)^2
\end{equation}
where the sum is performed on the 4 nearest neighbors of the atom and $\theta_i$ is the angle made by each triplet. $A$ to classify the atom as a solid-like ($A<0.4$) or a liquid-like ($A>0.4$) one. Consecutively, the structure under study is cut in slides, and the percentage of solid-like or liquid-like atoms is calculated in each slide. This order parameter is close to 1 in a crystal, around 0.2-0.4 in a liquid, and around 0.7-0.8 in an amorphous material. The order parameter as a function of $z$, the direction perpendicular to the disordered Si/c-Si, is represented for a simulation at 1273 K with the Stillinger-Weber potential (Fig. \ref{FigOPSW1273}) and at 1673 K with the Tersoff potential (Fig. \ref{FigOPT1673}). Fig. \ref{FigOPSW1273} clearly demonstrates that the disordered silicon for $z>30$\AA, while amorphous at the origin, transforms to a liquid during the initial equilibration stage. When the simulation of the annealing proceeds, the interface moves towards the right, as the recrystallization occurs, and the disordered part remains liquid. On the contrary with the Tersoff potential, the order parameter remains in the range $[$0.6-0.8$]$ during all the process of recrystallization. This is a first proof that the a-Si initially constructed is transformed into l-Si when the Stillinger-Weber potential is used, while it remains amorphous with the Tersoff potential.  

One could argue that these differences of behavior are due to the ability of the interatomic potentials to correctly predict the experimental melting temperature of a-Si. Indeed, the melting temperature of a-Si, $T_\mathrm{m}^a$, is measured in the range $[$1420-1530$]$ K\cite{Thompson84,Donovan83}. However, the Tersoff potential is known for its large overestimation of the melting temperature, for c-Si as for a-Si : Marqu\'es et al. \cite{Marques04} have determined a value of that the melting temperature of $T_\mathrm{m}^a$ = 2050 K, while for the Stillinger-Weber potential, a value of $T_\mathrm{m}^a$ = 1400 K was found by Brambilla et al. \cite{Brambilla00} and 1075 K by Albenze et al. \cite{SW115}. Indeed, Fig. \ref{FigOPSW(T)} shows that the temperature used with the Stillinger-Weber potential must be decreased to $[$473-673$]$ K to keep an order parameter of the amorphous kind after the initial equilibration stage. Therefore we think that not only the melting temperature enter into account but more globally the ability of the potential to handle or not an amorphous structure. The Stillinger-Weber potential clearly favors the liquid state, as already observed in the literature \cite{Broughton87}.

In order to further corroborate the transformation to a liquid state, the radial and angular distribution of the atoms have been calculated in the disordered layer after the equilibration stage. The results are respectively presented in Fig. \ref{FigRadDistrEqu} and \ref{FigAngDistrEqu}. After a simulation with the Tersoff and SW115 potentials, the typical characteristics of an amorphous material are preserved: distinct peak of the second neighbours with a gap between first and second neighbours, and angular distribution comparable to the initial one (WWW a-Si). With Stillinger-Weber, EDIP and Lenosky on the other hand, there is no gap between the peaks of the first and second neighbors in the radial distribution (see Fig. \ref{FigRadDistrEqu}), these second peak being even much less pronounced. Moreover, the angular distribution is drastically smeared out (see Fig. \ref{FigAngDistrEqu}). All such structural features are far more typical of a liquid \cite{Waseda75,Stich89}.

Consequently, our understanding of MD simulations of the recrystallization of a-Si using empirical potentials is the following. The Tersoff and SW115 potentials are adapted to the description of the SPE. The amorphous character is preserved with these potentials and the velocity of recrystallization follows an Arrhenius law as observed experimentally. The SW115 potential gives however a recrystallization velocity two orders of magnitude faster than the experiment. The Tersoff potential gives an excellent agreement with the law of Olson and Roth\cite{Olson88}. Its drawback is its excessive value of the melting temperature (for a-Si as for c-Si). We have tried to apply a simple rescaling of the temperature $T_\mathrm{rescaled} = T_\mathrm{simu} \times T_\mathrm{m,exp}^c/T_\mathrm{m,simu}^c$, $T_\mathrm{simu} \times 1685/2400$ in the case of the Tersoff potential. This approach is not satisfying since the simulation has moved away from the experiment by nearly 4 order of magnitudes as shown in Fig. \ref{FigTempNorm}.

The Stillinger-Weber, EDIP and Lenosky potentials are not adapted to the simulation of the SPE of a-Si. These three potentials seem to have a tendency to transform the amorphous structure into a liquid one, even below the nominal bulk melting temperature. Nevertheless, the resort to the Stillinger-Weber potential is perfectly justified to simulate the LPE, the experimental results obtained by pulse laser annealing \cite{Galvin85} being in the margin of incertitude of our simulations. The sluggish behavior of the EDIP potential, already mentioned by Albenze and Clancy \cite{Albenze05}, is due to its formulation developed to handle non-tetrahedral environments. This environment dependency is a disadvantage for this particular problem, the return to the crystal arrangement is not enough enforced. Finally, the Lenosky potential is also a good candidate for the simulation of LPE if the temperature is scaled to account for the differences in metling temperatures (see Fig. \ref{FigTempNorm}). 

\section{Conclusion}

The recrystallization of an amorphous silicon layer on a crystalline silicon substrate has been simulated by means of molecular dynamics. The initial a-Si on c-Si stack has been constructed by the WWW method and the structural properties of the resulting amorphous were carefully analyzed to validate the atomic arrangement in the structure with respect to the available experimental results. Subsequently, the annealing of this amorphous-crystal stack was studied using five of the most widely used interatomic potentials for silicon. A systematic campaign of molecular dynamics simulations as a function of temperature in the appropriate interval to observe amorphous recrystallization was carried out to extract the velocity of recrystallization versus the temperature.  A wide disparity is obtained as a function of the interatomic potential, underscoring the limitations of the different empirical potentials that are not able to treat with an equivalent and consistent degree of accuracy the three Si phases of interest in the technological process (diamond crystal, amorphous and liquid). A similar shortcoming had been already pointed out by Ding and Andersen \cite{Ding86} for the case of germanium. However, the results can be classified in two categories. The first category is the expected solid phase epitaxy of the a-Si, obtained with the Tersoff and SW115 potentials. The second category corresponds rather to a liquid phase epitaxy as obtained with the potentials of Stillinger-Weber, EDIP or Lenosky, due to the transformation of the starting a-Si cluster into a liquid one. The reliability of the empirical potential molecular dynamics compared to experiments is shown to be very good, provided that a suitable potential is chosen to investigate each different phenomenology. Based on  the results of the present study, the best candidates turn out to be Tersoff for the SPE and Stillinger-Weber for the LPE. 

\section{Acknowledgements} 
This work is part of the common laboratory STMicroelectronics/IEMN supported by the MINEFI. The numerous computer simulations would not have been possible without the reliable park of workstations managed by Jean-Michel Droulez. The authors thank W. Windl for the suggestion to look at the Lenosky potential and Guy Allan for  discussion about the amorphous cluster.

\begin{table}
\begin{tabular}{|c|c|c|c|c|c|c|c|c|}
\hline
\multicolumn{3}{|c|}{Conditions}&Coordination Number&Twofold&Threefold&Fourfold&Fivefold&Sixfold\\
\hline\hline
\multicolumn{3}{|c|}{Initial Interface}&3.73&10.6\%&16.9\%&55.7\%&16.8\%&0\%\\
\hline\hline
Duration&Temperature&Potential&&&&&&\\
 4ps&1273K&Tersoff&3.92&0.6\%&10.7\%&84.20\%&4.50\%&0\%\\
 4ps&1273K&SW115&3.72&4.3\%&18.50\%&77.20\%&0\%&0\%\\
 4ps&1273K&SW&3.86&0\%&15.4\%&79\%&5.6\%&0\%\\
 4ps&1273K&EDIP&4.08&0\%&3.0\%&85.3\%&11.7\%&0\%\\
 4ps&1273K&Lenosky&4.63&0\%&6.4\%&49.7\%&23.1\%&20.8\%\\
\hline\hline
\end{tabular}
\caption{Distribution of the mean coordination number for the initial a-Si/c-Si interface. The evolution of this distribution during the equilibration step of4 ps with the various interatomic potential is also represented for the same  condtions (velocity scaling, 1273K). The equilibration step improves the quality of the interface since the fraction of fourfold silicon atoms is obviously enhanced. On the other side, the choice of the interatomic potential has a clear influence on the defects at the interface even for a very short annealing.}
\label{table:interface_coordination}
\end{table}

\begin{figure}[htb]
\includegraphics[scale=0.7]{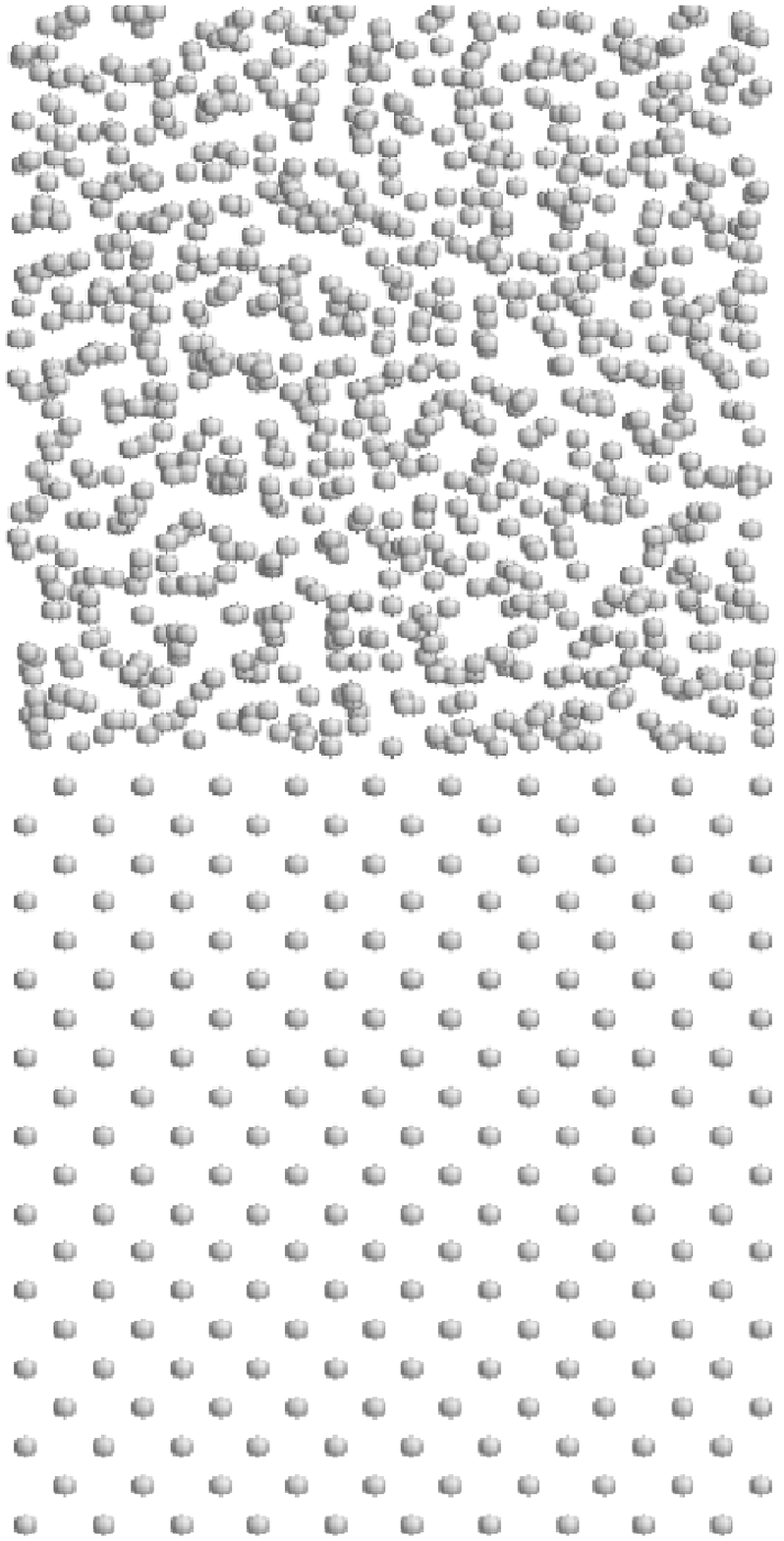}
\caption{a-Si on c-Si stacks (1905 atoms).
\label{FigaSioncSi}}
\end{figure}

\begin{figure}[htb]
\includegraphics[scale=0.9]{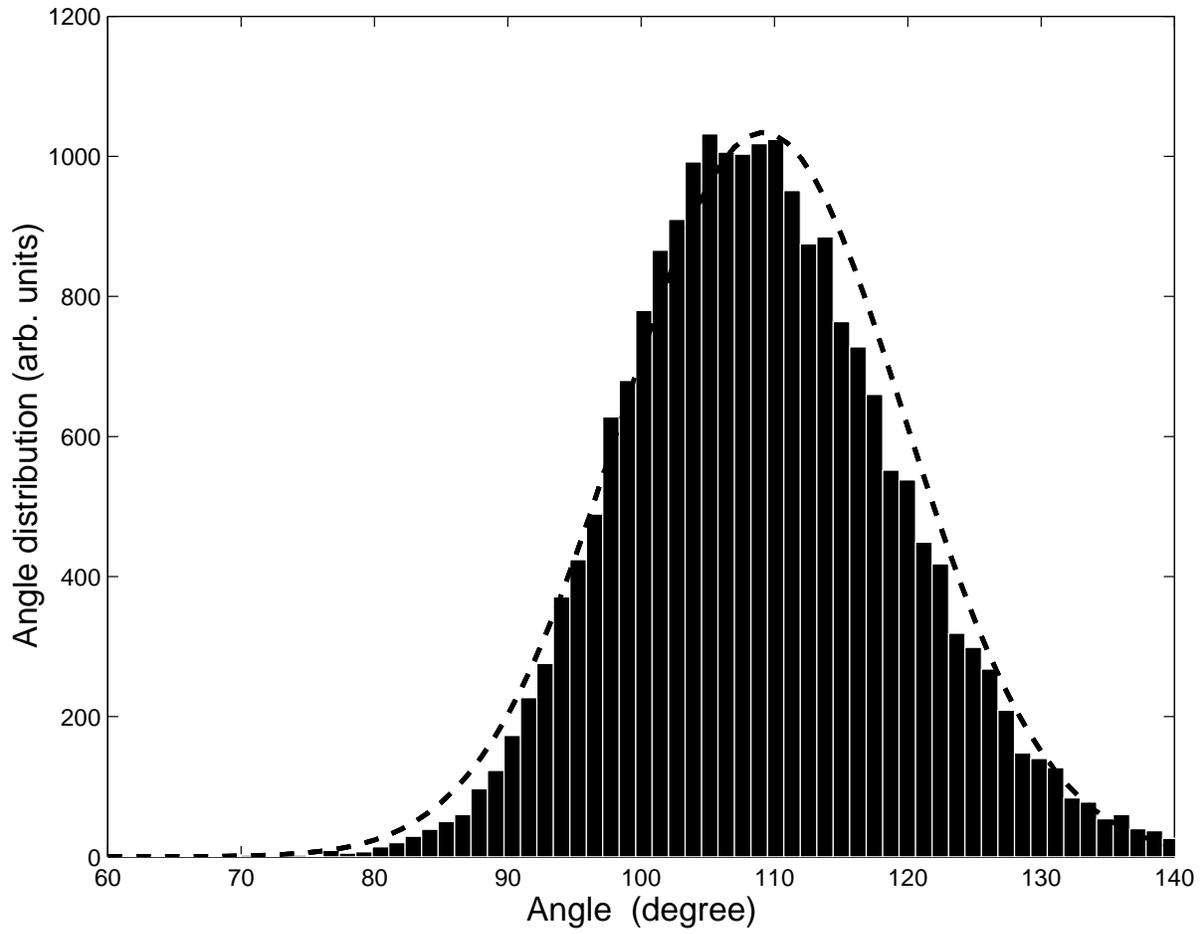}
\caption{Angular distribution in the a-Si cluster.
\label{FigAngDist}}
\end{figure}

\begin{figure}[htb]
\includegraphics[scale=0.9]{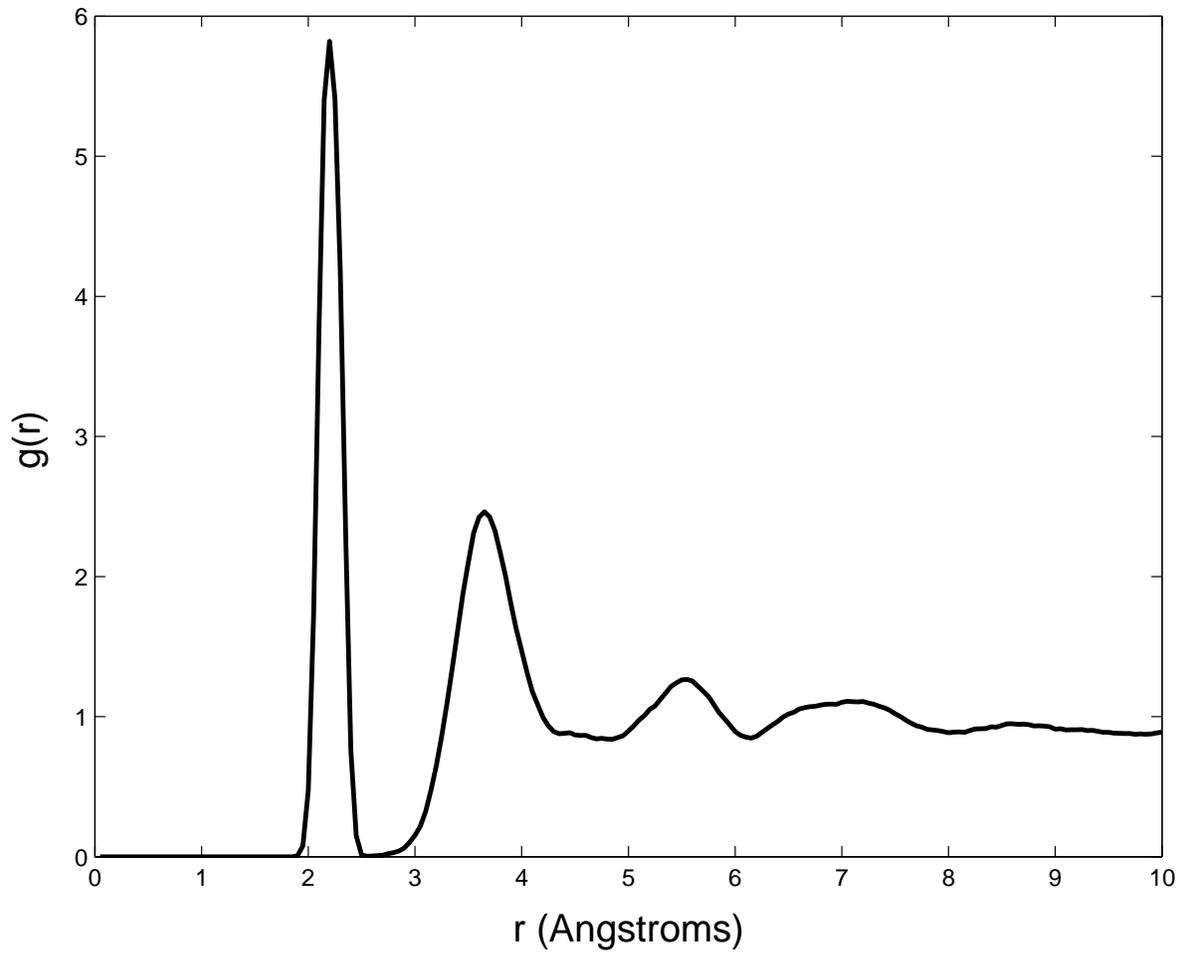}
\caption{Radial distribution in the a-Si cluster.
\label{FigRadDist}}
\end{figure}

\begin{figure}[htb]
\includegraphics[scale=0.9]{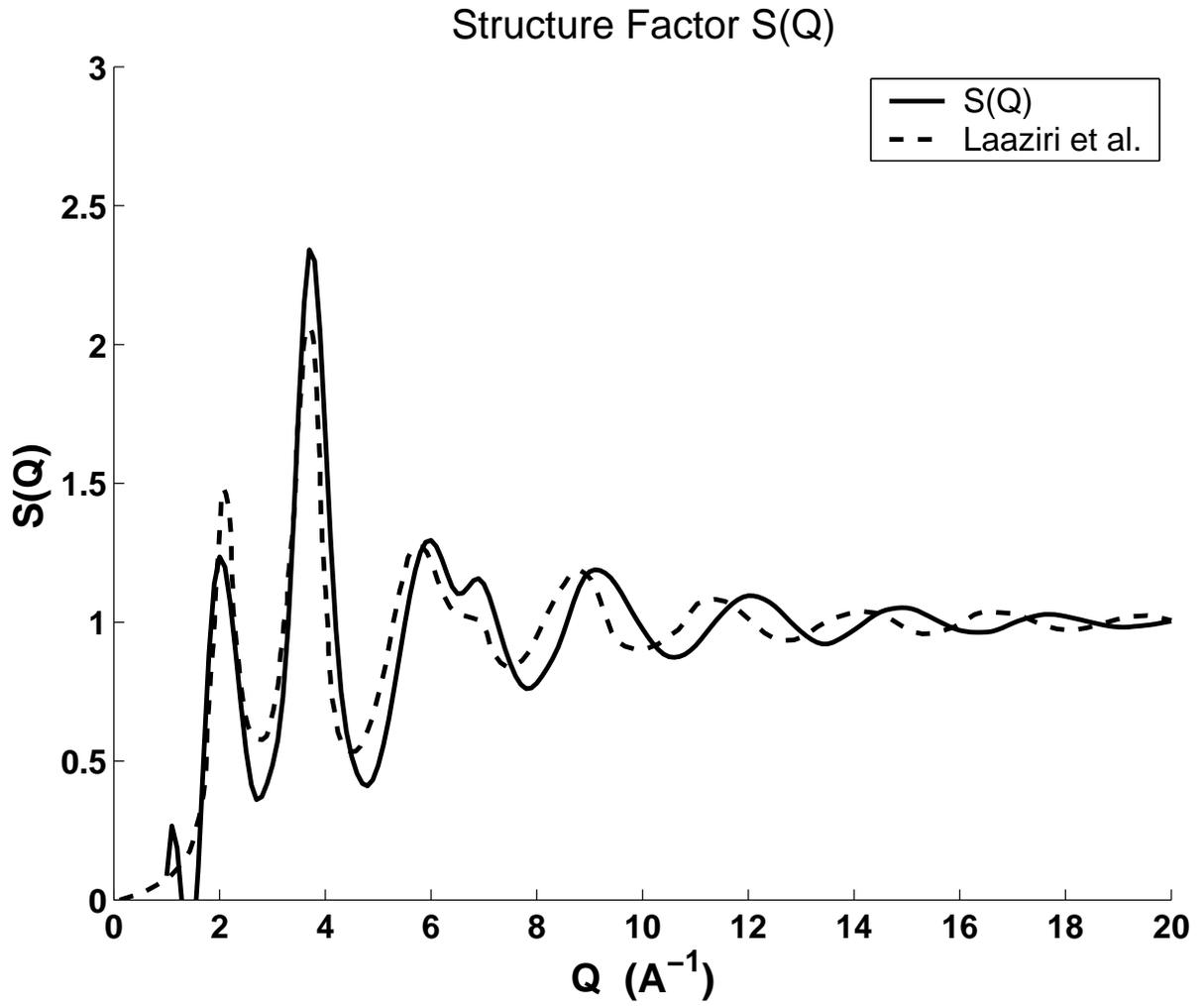}
\caption{Comparison between the  structure factor of the WWW a-Si cluster and the experiment from Laaziri {\it et al.} \cite{Laaziri99}.
\label{FigFactStruc}}
\end{figure}

\begin{figure}[htb]
\includegraphics[scale=2]{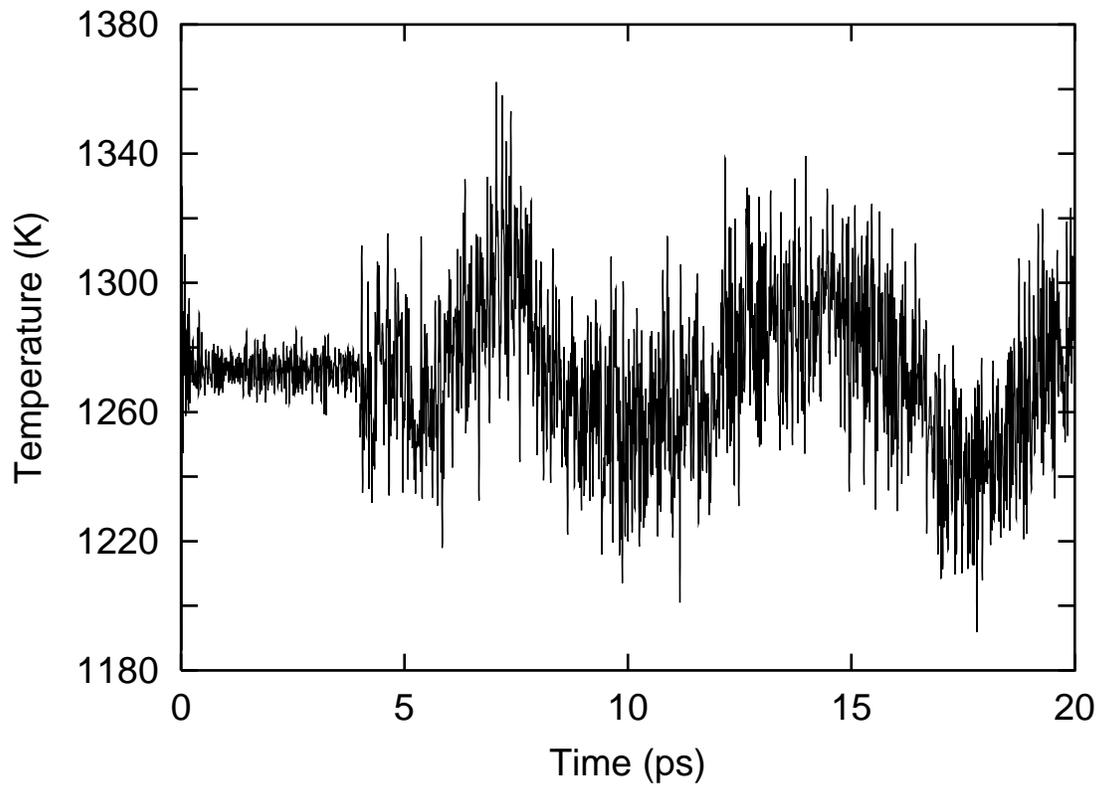}
\caption{Temperature versus time for a command of 1273 K. The velocities are initiated using a Maxwell-Boltzman distribution. From 0 to 4 ps, the phase of equilibration of the simulation, a scaling of the velocities is used to follow the command of 1273 K. After 4 ps, the phase of acquisition begins thanks to the use of a Nos\'e-Hoover thermostat. 
\label{FigTemp}}
\end{figure}

\begin{figure}[htb]
\centering
\includegraphics[scale=0.9]{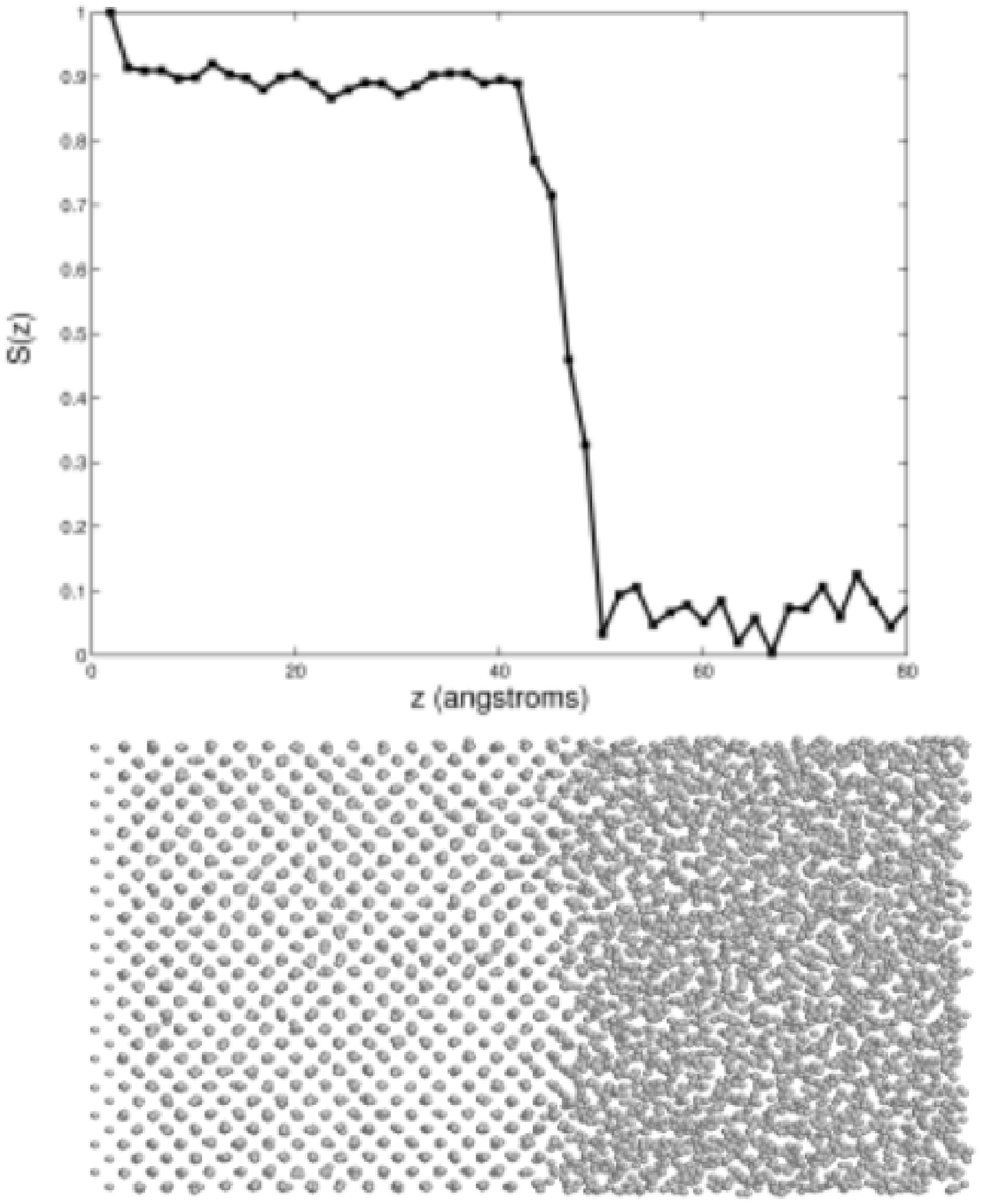}
\caption{Structure factor along the growth direction calculated for a stack of 8034 atoms. The structure factor and a representation of the atoms are aligned to evidence the link between the order/disorder and the value of the structure factor.
\label{FigSz}}
\end{figure}

\begin{figure}[htb]
\includegraphics[scale=0.9]{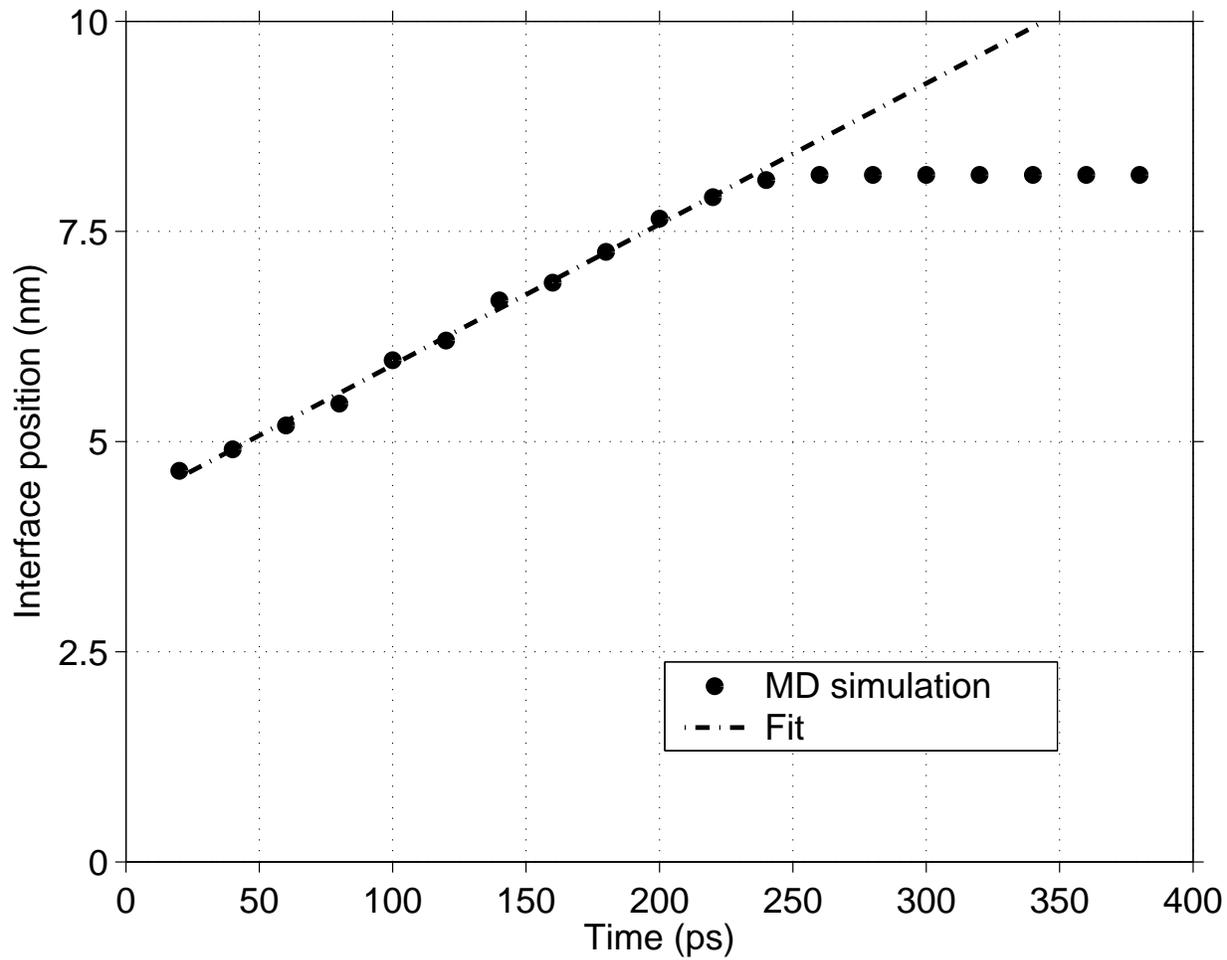}
\caption{a-Si/c-Si interface position versus the annealing time and fit of the linear region to extract the velocity of recrystallization. 
\label{FigFitInterface}}
\end{figure}

\begin{figure}[htb]
\includegraphics[scale=0.9]{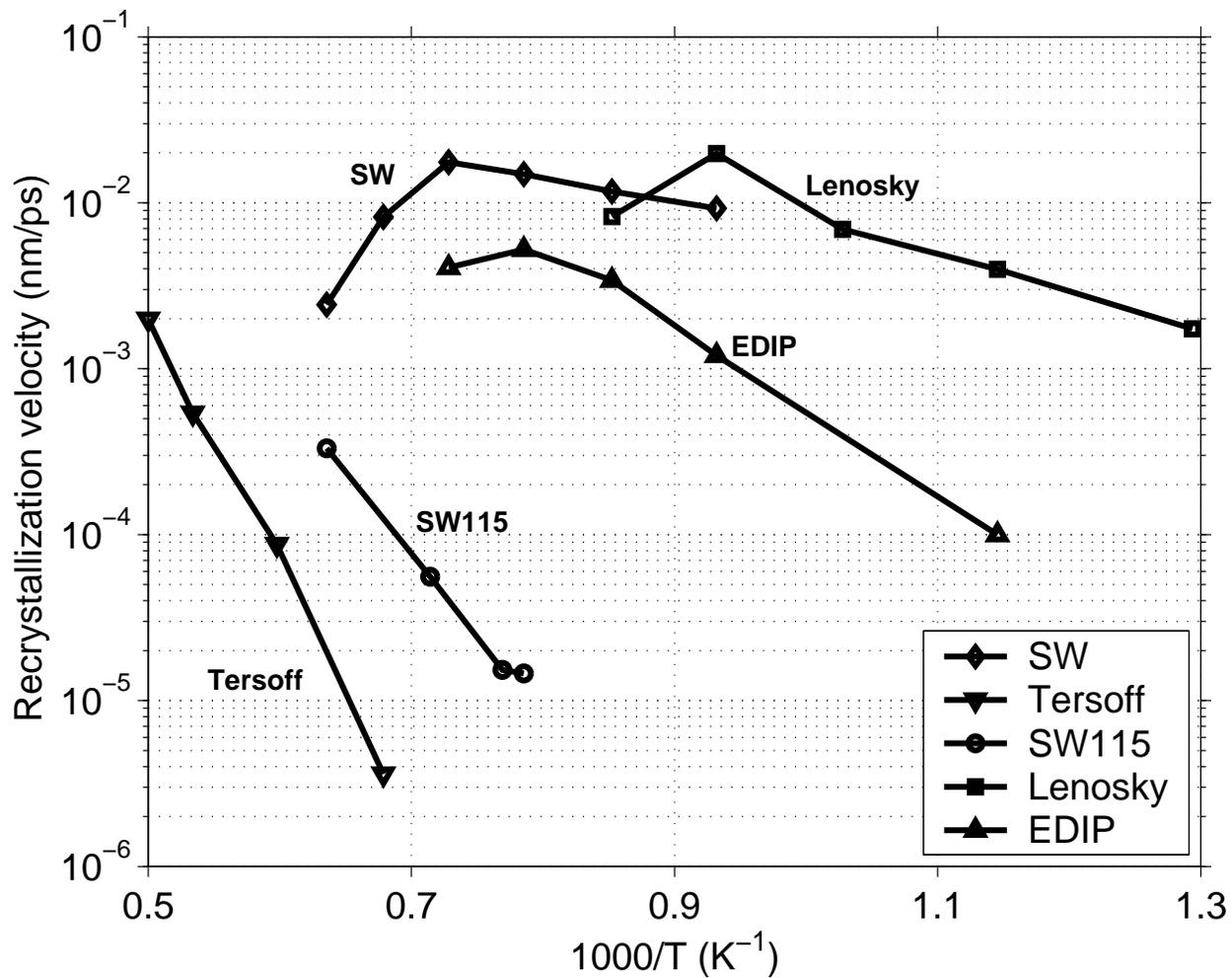}
\caption{ Recrystallization velocity versus the inverse temperature for the 5 interatomic potentials.
\label{FigVelCalc}}
\end{figure}

\begin{figure}[htb]
\includegraphics[scale=0.9]{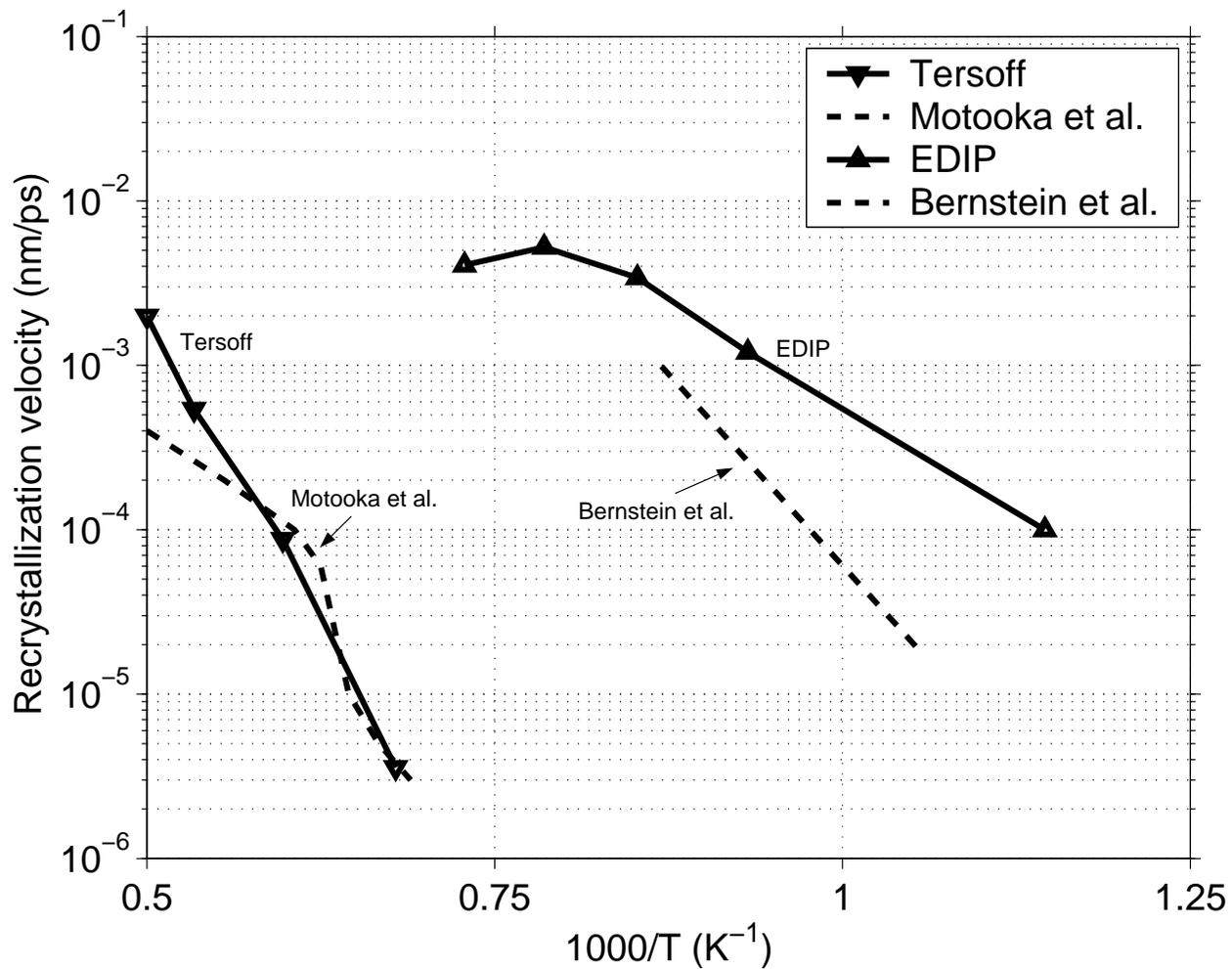}
\caption{Recrystallization velocity versus the inverse temperature. Comparison of our simulations using the Tersoff and EDIP potentials with the works of Motooka et al. \cite{Motooka00} and Bernstein et al. \cite{Bernstein00}.
\label{FigCompSimu}}
\end{figure}

\begin{figure}[htb]
\includegraphics[scale=0.9]{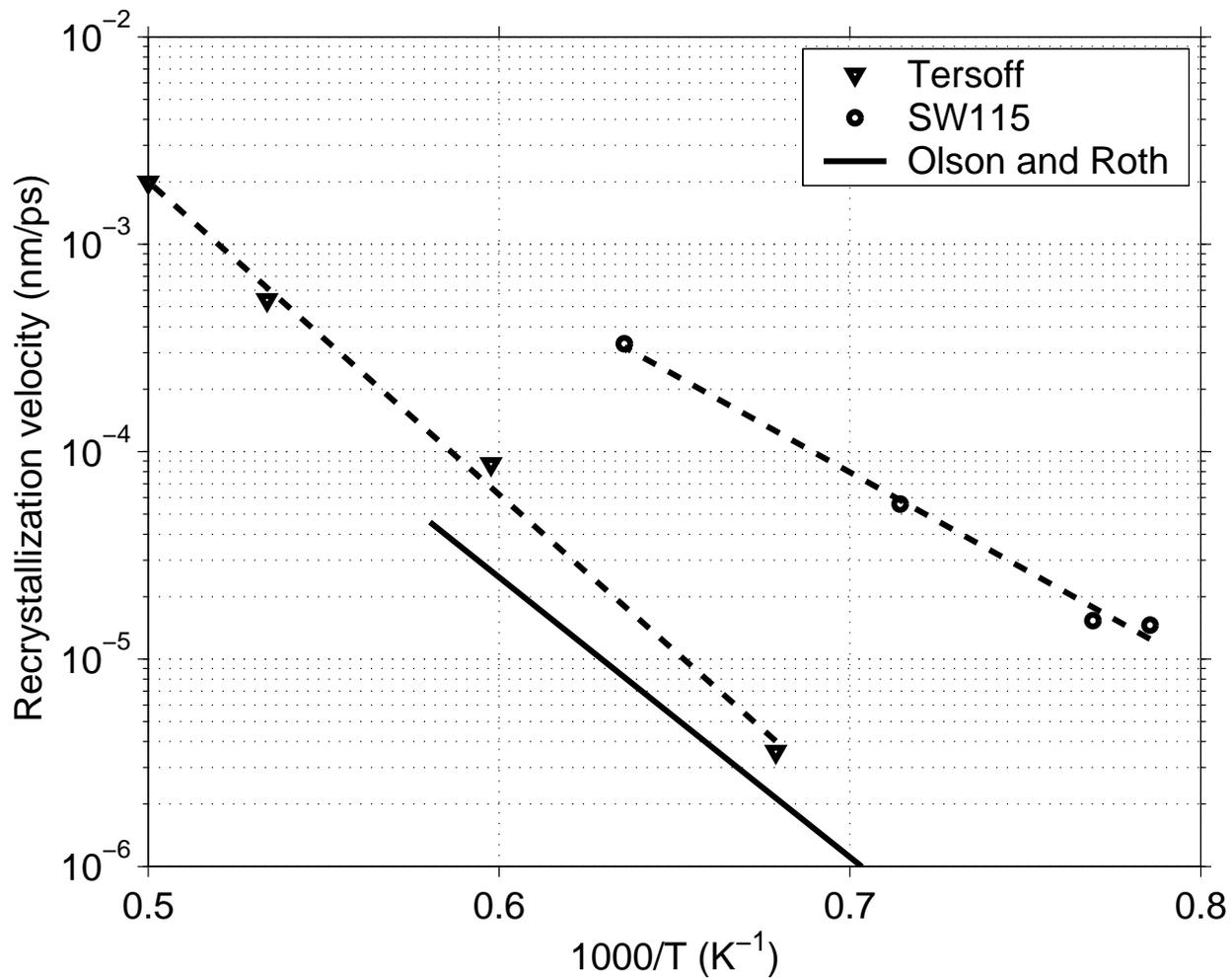}
\caption{Simulations of the recrystallization velocities with the Tersoff and SW115 potentials (points), their adjustment by Arrhenius laws (dashed lines), and the experimental results of Olson and Roth \cite{Olson88} (continuous line).
\label{FigFitTersoffSW115}}
\end{figure}

\begin{figure}[htb]
\includegraphics[scale=2]{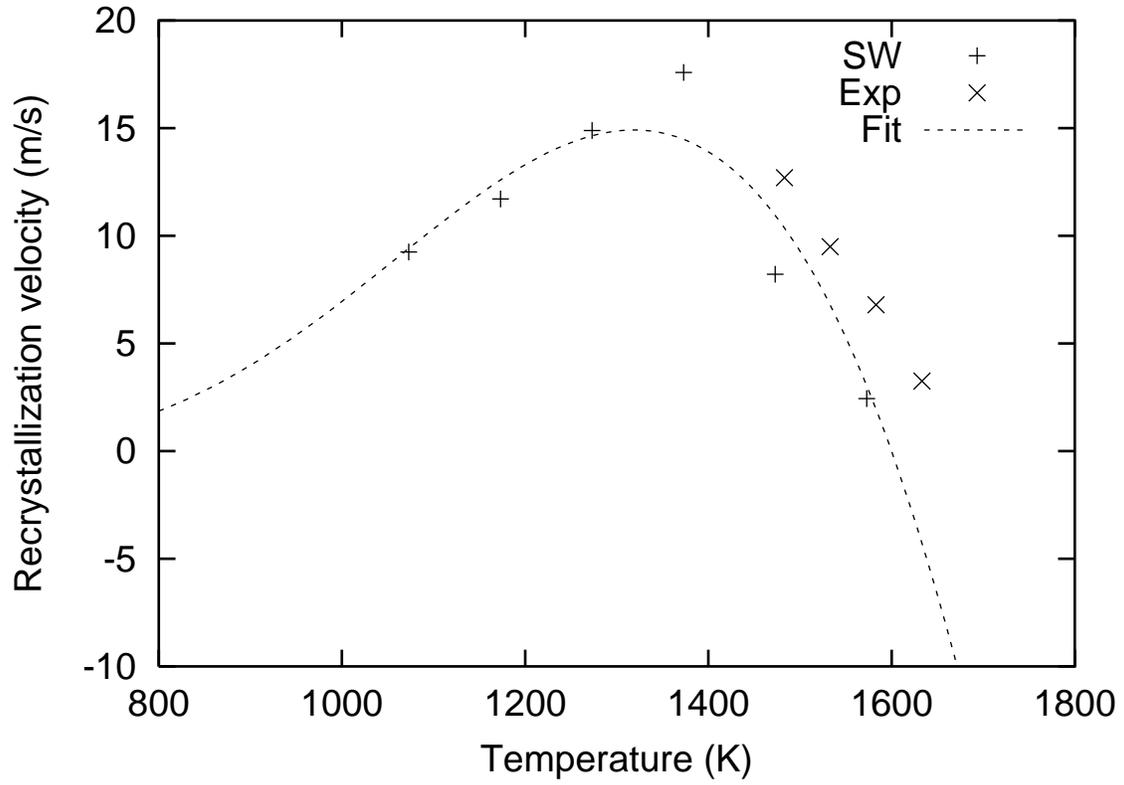}
\caption{Simulations of the recrystallization velocities with the Stillinger-Weber potential ($+$), measurement of the velocity of recrystallization of melted Si \cite{Galvin85} ($\times$) and fit of the simulations according to Eq. \ref{EqVelLiqu}.
\label{FigSWExpFit}}
\end{figure} 

\begin{figure}[htb]
\includegraphics[scale=2]{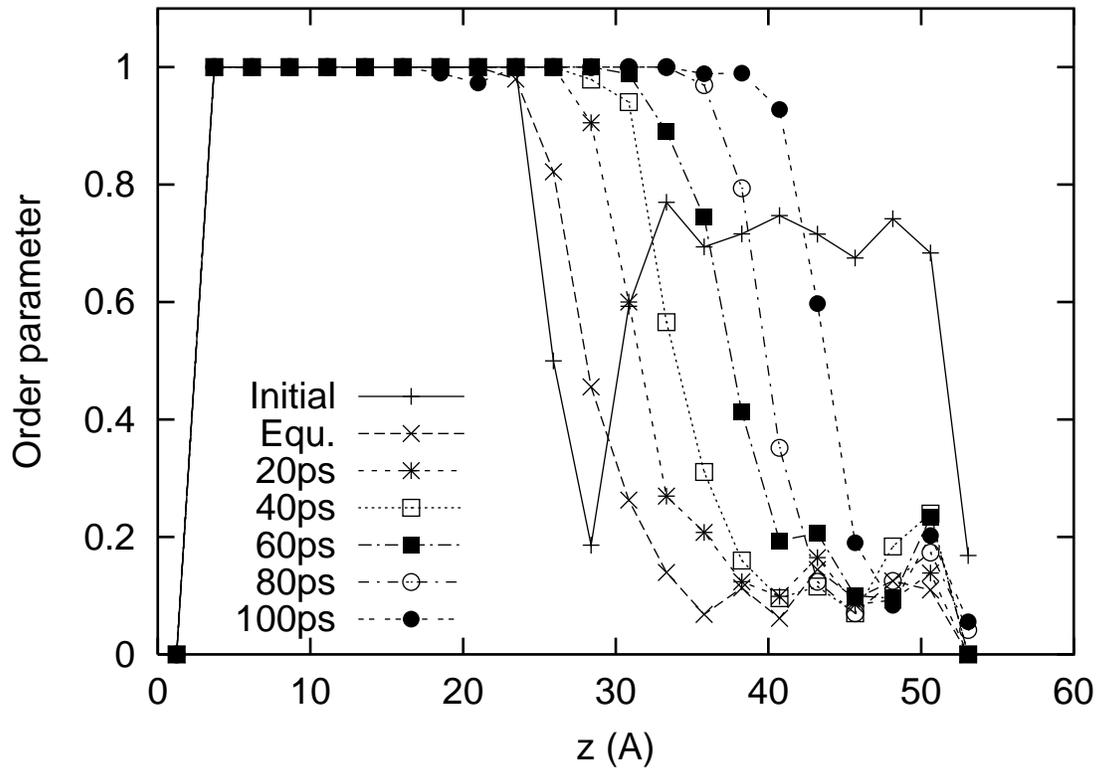}
\caption{Order parameter in the ``disordered"/c-Si stack initially constructed (Initial), after the stage of equilibration (Equ.) and an annealing during 20, 40, 60, 80 and 100 ps at 1273 K. Simulation with the Stillinger-Weber potential.
\label{FigOPSW1273}}
\end{figure} 

\begin{figure}[htb]
\includegraphics[scale=2]{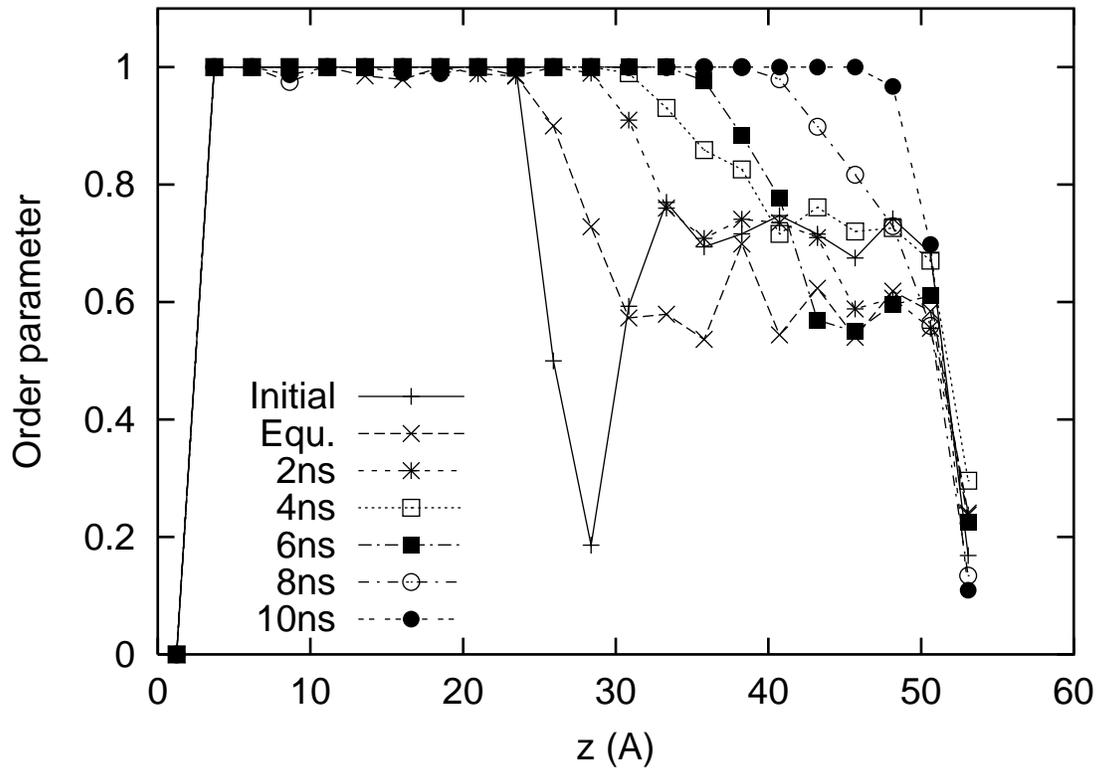}
\caption{Order parameter in the ``disordered"/c-Si stack initially constructed (Initial), after the stage of equilibration (Equ.) and an annealing during 2, 4, 6, 8 and 10 ns at 1673 K. Simulation with the Tersoff potential.
\label{FigOPT1673}}
\end{figure} 

\begin{figure}[htb]
\includegraphics[scale=2]{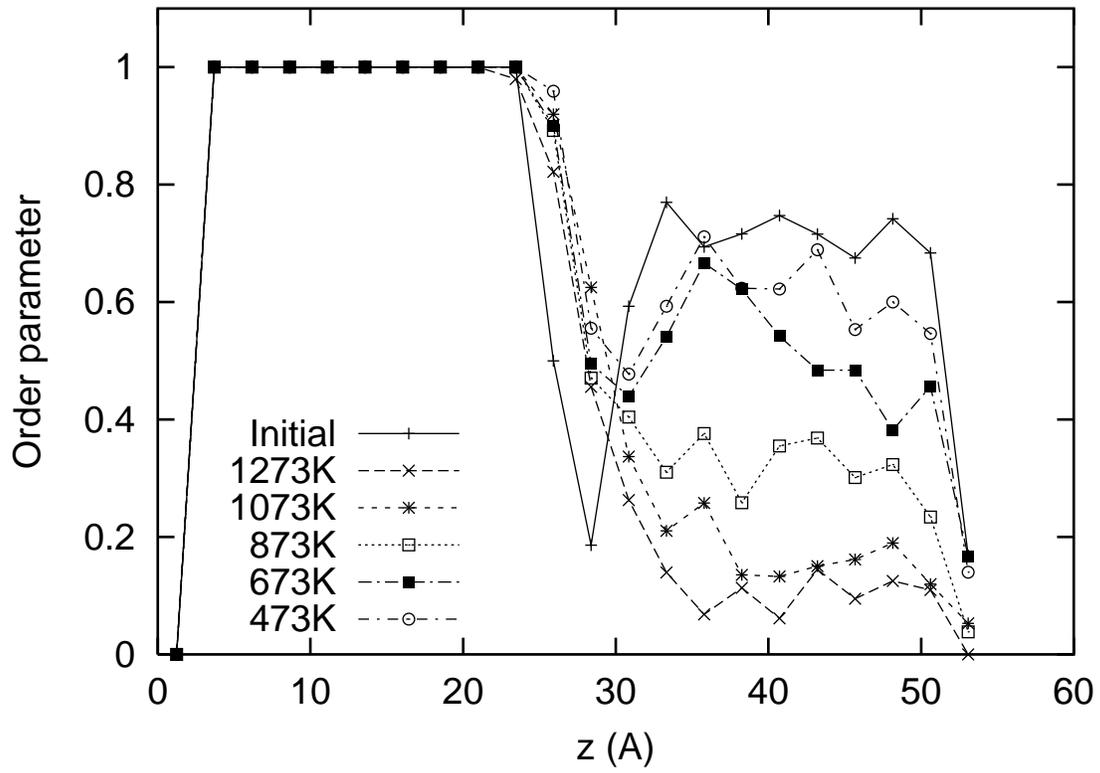}
\caption{Order parameter in the ``disordered"/c-Si stack initially constructed (Initial), and after the stage of equilibration at 473, 673, 873, 1073 and 1273 K simulated with the Stillinger-Weber potential.
\label{FigOPSW(T)}}
\end{figure} 

\begin{figure}[htb]
\includegraphics[scale=0.8]{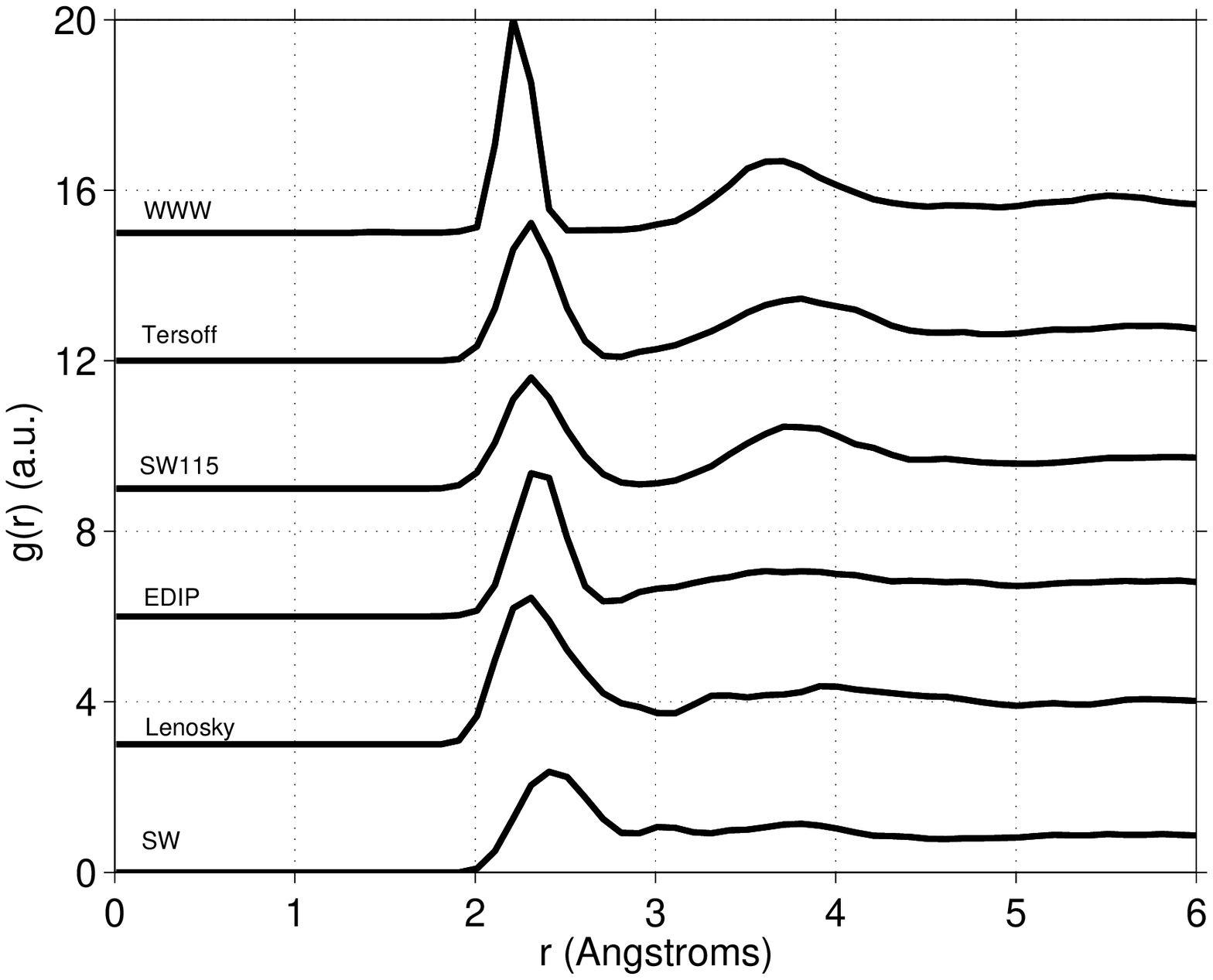}
\caption{Radial distribution in the disordered part. Initial WWW system (WWW), and after equilibration with the Tersoff, SW115, Stillinger-Weber (SW), EDIP and Lenosky potentials. 
\label{FigRadDistrEqu}}
\end{figure} 

\begin{figure}[htb]
\includegraphics[scale=0.9]{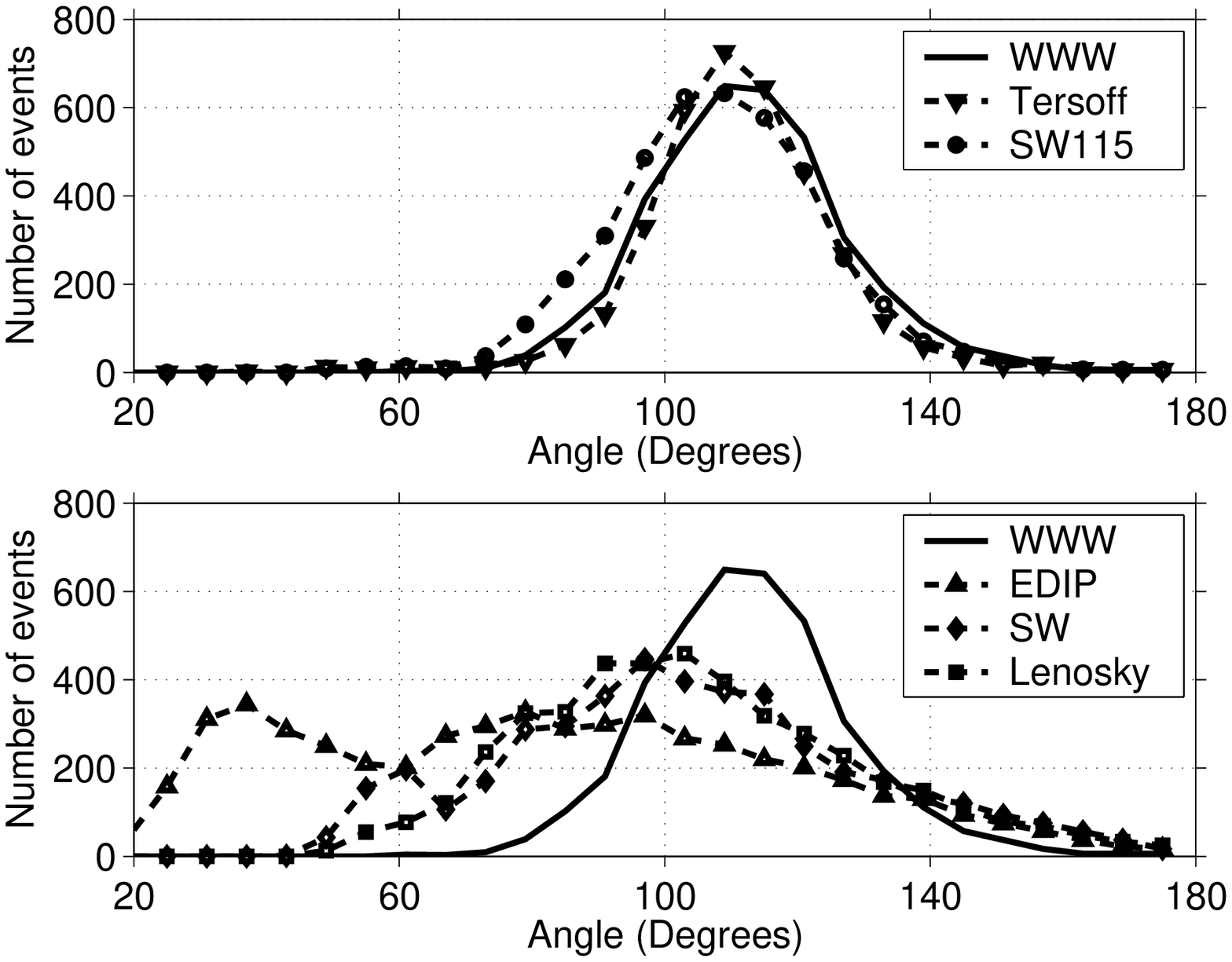}
\caption{Angular distribution in the disordered part. Initial WWW system (WWW), and after equilibration with the Tersoff, SW115, Stillinger-Weber (SW), EDIP and Lenosky potentials.
\label{FigAngDistrEqu}}
\end{figure}

\begin{figure}[htb]
\includegraphics[scale=0.8]{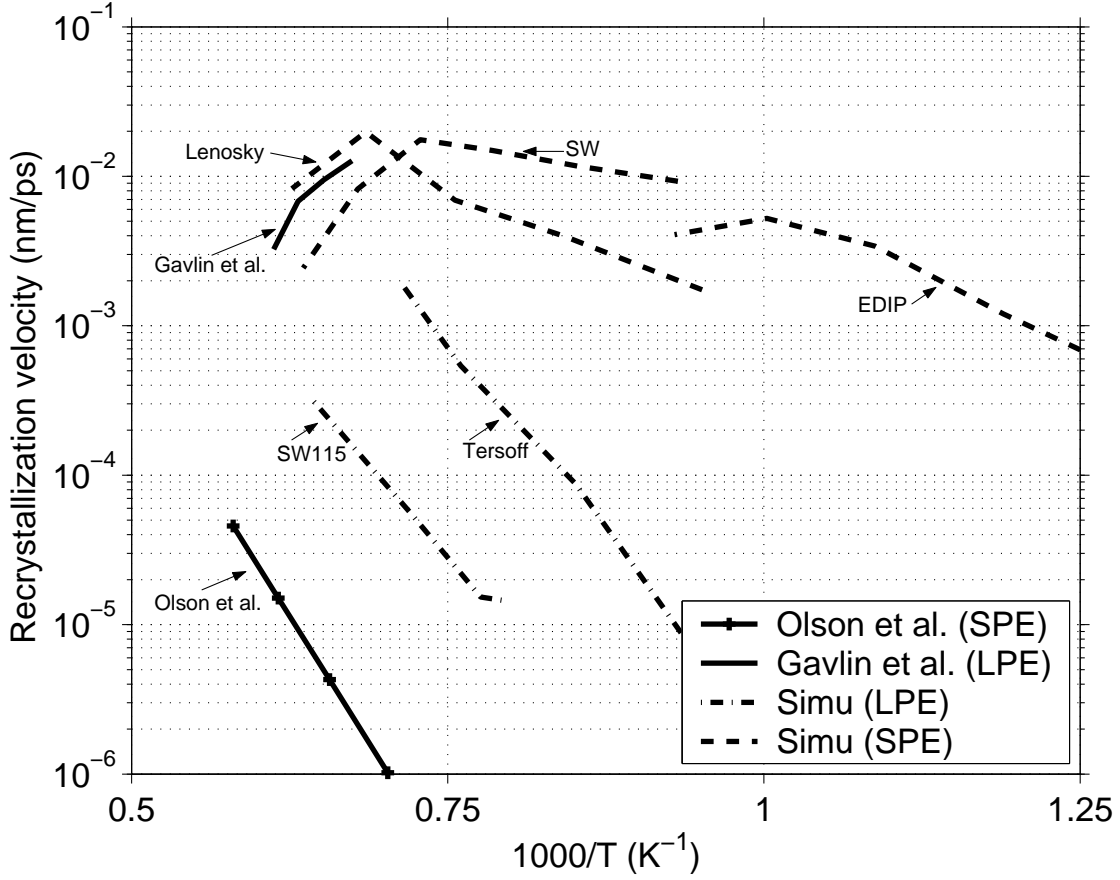}
\caption{ Experimental and calculated velocities of recrystallization versus the temperature. The temperature has been rescaled for the simulations following $T_\mathrm{rescaled} = T_\mathrm{simu} \times T_\mathrm{m,exp}^c/T_\mathrm{m,simu}^c$, where the experimental melting temperature of c-Si $T_\mathrm{m,exp}^c$ is equal to 1685 K and the simulated melting temperature of c-Si $T_\mathrm{m,simu}^c$ is potential dependent. \vspace{7cm}
\label{FigTempNorm}}
\end{figure}
\newpage

\end{document}